\documentclass[%
aip,%
amsmath,amssymb,floatfix,
reprint,%
 longbibliography,
reprint,%
author-year,%
author-numerical,%
]{revtex4-1}

 \usepackage{type1ec} 

\usepackage{amsmath,amsthm,latexsym,amssymb,amsfonts,epsfig}

\usepackage[english]{babel}
\usepackage[utf8]{inputenc}			
\usepackage{graphicx, texdraw}   
\usepackage{bm}

\usepackage{mathrsfs}

\usepackage{mathtools}  
\usepackage{makecell}   

\usepackage[left= 1.3cm, right = 1.3cm, top= 2.28cm, bottom = 2.3cm]{geometry}

\usepackage{enumerate}

\usepackage[colorlinks=true,citecolor=OliveGreen, linkcolor=blue]{hyperref}

\usepackage{csquotes} 

\usepackage[usenames, dvipsnames]{color}

\usepackage{amsmath}
\usepackage{amsfonts}
\usepackage{amssymb}
\usepackage{braket}

\usepackage{float}
\floatstyle{plaintop}
\restylefloat{table}

\usepackage{bigints}

\newcommand{\vect}[1]{\boldsymbol{#1}}

\newcommand{\bNabla}{\boldsymbol{\nabla}}

\newcommand{\R}{\vect{r}}

\newcommand{\kS}{\kappa_\mathrm{S}}

\newcommand{\kB}{\kappa_\mathrm{B}}

\newcommand{\alphaB}{\alpha_\mathrm{B}}

\newcommand{\Intd}{\mathrm{d}}

\newcommand{\betaB}{\beta_{\mathrm{B}}}

\newcommand{\bigO}{\mathcal{O}}

\newcommand{\halb}{\frac{1}{2}}
\newcommand{\threehalb}{\frac{3}{2}}
\newcommand{\fivehalb}{\frac{5}{2}}

\allowdisplaybreaks

\begin{document}
\title{Asymmetric Stokes Flow Induced by a Transverse Point Force Acting Near a Finite-Sized Elastic Membrane}

\author{Abdallah Daddi-Moussa-Ider}
\email{abdallah.daddi.moussa.ider@uni-duesseldorf.de}
\affiliation
{Institut f\"{u}r Theoretische Physik II: Weiche Materie, Heinrich-Heine-Universit\"{a}t D\"{u}sseldorf, Universit\"{a}tsstra\ss e 1, D-40225 D\"{u}sseldorf, Germany}

\date{\today}

\begin{abstract}

	A deep understanding of the physical interactions between nanoparticles and target cell membranes is important in designing efficient nanocarrier systems for drug delivery applications.
	Here, we present a theoretical framework to describe the hydrodynamic flow field induced by a point-force singularity (Stokeslet) directed parallel to a finite-sized elastic membrane endowed with shear and bending rigidities. 
	We formulate the elastohydrodynamic problem as a mixed-boundary-value problem, which we then reduce into a well-behaved system of integro-differential equations.
	It follows that shear and bending linearly decouple so that the solution of the overall flow problem can be obtained by linear superposition of the contributions arising from these modes of deformation.
	Additionally, we probe the effect of the membrane on the hydrodynamic drag acting on a nearby particle, finding that, in a certain range of parameters, translational motion near an elastic membrane with only energetic resistance toward shear can, surprisingly, be sped up compared to bulk fluid.
	Our results may find applications in microrheological characterizations of colloidal systems near elastic confinements. 	

\end{abstract}
\maketitle

\section{Introduction}

Hydrodynamic interactions in confined geometries~\cite{diamant09} are of pivotal importance in a variety of biological and physiological processes ranging from the transport of cells and macromolecules in stenosed arterial walls of blood vessels~\cite{muller14, fedosov2012margination, cilla14, bozsak14, karner01, ai06, kaoui2018computer} to the foraging behavior of commensal bacteria in human and animal intestine~\cite{berry13, backhed05, farhadi03}.
In addition, surface-related effects on the transport behavior of particulate flows play a crucial role in many biomedical and pharmaceutical applications.
A prime example is given by drug delivery of nanocarriers to organelles~\cite{langer98, DeJong2008, veiseh10, rosenholm10, colson12, maeda13, liu16} prior to their uptake by cell membranes via endocytosis~\cite{hillaireau09, Doherty_2009, oh14, agudo16, ohta2020brownian}.
In these scenarios, nanoparticles frequently enter the close vicinity of confining elastic interfaces, which is known to drastically alter their behavior, dynamics, and energetics in viscous media.

On the micron scale, fluid flows are characterized by low Reynolds numbers, $\operatorname{Re} = \rho L V / \eta \ll 1$, where~$L$ and~$V$ are, respectively, typical length and velocity scales of the flow, $\eta$~denotes the dynamic viscosity of the surrounding fluid and $\rho$ the density.
Accordingly, viscous forces dominate inertial forces.
Under these conditions, the fluid dynamics is well described by the linear Stokes equations~\cite{happel12}.
Over the last couple of decades, there have been tremendous research efforts in addressing the behavior of hydrodynamically interacting particles near interfaces including a planar rigid wall bounding a semi-infinite fluid medium~\cite{mackay61, gotoh82, cichocki98, lauga05, swan07, franosch09, felderhof12, decorato15, huang15, rallabandi17obstacle, kaoui09hb}, an interface separating two immiscible Newtonian fluids~\cite{lee79, berdan81, blawz10theory, blawz10, krueger13}, a rough surface~\cite{kurzthaler2020particle}, or a deformable membrane~\cite{felderhof06, bickel06, bickel07, shlomovitz14, boatwright14, daddi16, daddi16b, daddi16c, junger15, junger16, daddi18coupling, daddi2019frequency}.
The latter type of interface stands apart as it endows the system with memory effects owing to the elastic nature.

In a preceding paper~\cite{daddi19jpsj}, we investigated theoretically the axisymmetric Stokes flow induced by a point-force singularity (Stokeslet) directed \textit{normal} to a finite-sized elastic disk (membrane) featuring resistance toward shear and bending.
In this contribution, we complement and extend these results by providing the solution of the asymmetric flow problem for a transversely directed point force acting \textit{parallel} to the surface of the disk.
Still, the point force remains located on the center axis of the undeformed disk.
The solution of the elastohydrodynamic problem is likewise formulated as a classic mixed-boundary-value problem~\cite{sneddon66} which is subsequently reduced into a system of dual integral equations~\cite{polyanin08}.
For their solutions, we employ the well-established methods introduced by Sneddon~\cite{sneddon60} and Copson~\cite{copson61} to yield a system of integro-differential equations amenable to numerical integration. 
This solution technique has previously been employed to determine the flow field induced by a Stokeslet acting near a hard disk~\cite{kim83, daddi20pof} or between two coaxially positioned rigid disks~\cite{daddi20jfm}.
For infinite shear and bending rigidities, we provide exact analytical solutions of the resulting integral equations for the induced flow field.

The remainder of the paper is organized as follows. 
In Sec.~\ref{sec:math}, we formulate the asymmetric flow problem for a transversely directed Stokeslet and introduce a model for the elastic membrane based on the Skalak and Helfrich models for the descriptions of shear and bending deformation modes, respectively.
In Sec.~\ref{sec:loesung}, we formulate the mixed-boundary-value problem and express its solution in terms of a system of integro-differential equations.
Thereupon, we compute in Sec.~\ref{sec:mobilitaet} the hydrodynamic mobility function for a point-like particle asymmetrically moving close to a finite-sized elastic membrane, and we compare our solution with that obtained near a hard, undeformable disk of the same size.
Finally, concluding remarks are contained in Sec.~\ref{sec:abschluss}.

\section{Mathematical formulation}
\label{sec:math}

\subsection{Stokes hydrodynamics}

\begin{figure}	
	\centering
	\includegraphics[scale=1.15]{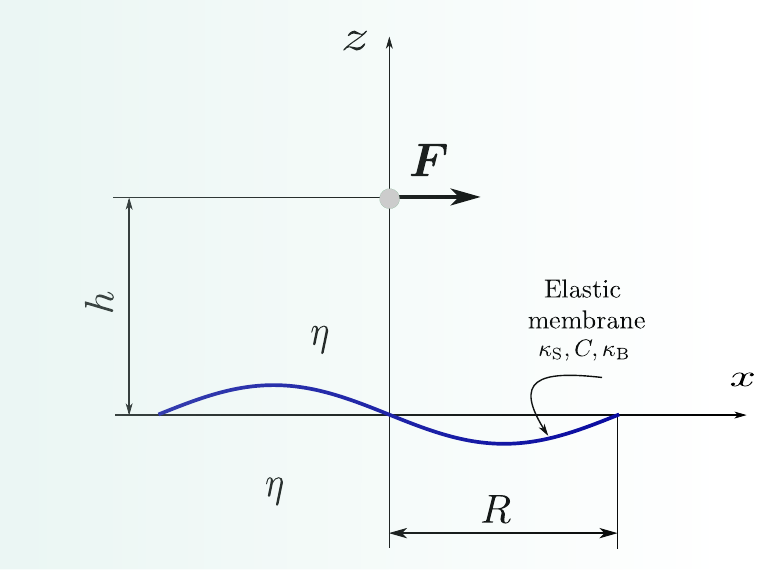}
	\caption{(Color online) Schematic illustration of a cross-section of the system under investigation. 
	A point-force singularity (Stokeslet) is directed tangent to the surface of a finite-sized deformable piece of an infinitely thin circular membrane of radius~$R$.
	This Stokeslet is located at a distance~$h$ along the center axis of the membrane above it.
	The elastic membrane is endowed with resistance toward shear and bending. 
	We denote by $\eta$ the dynamic viscosity of the surrounding incompressible fluid.}	
	\label{fig1}
\end{figure}

We examine the low-Reynolds-number flow induced by a point-force singularity acting tangent to an initially flat finite-sized elastic membrane of radius~$R$.
For this purpose, we adopt a frame of reference, the origin of which coincides with the center of the initially undeformed membrane of circular circumference, as schematically illustrated in Fig.~\ref{fig1}.
At low Reynolds numbers, the fluid dynamics is governed by the forced Stokes equations~\cite{happel12}
\begin{subequations}\label{StokesGleischungen}
	\begin{align}
		-\boldsymbol{\nabla}p + \eta \boldsymbol{\nabla}^2 \vect{v} 
		+ \vect{F} \delta \left( \R-\R_0 \right) &= \vect{0} \, , \\
		\boldsymbol{\nabla} \cdot \vect{v} &= 0 \, ,     
	\end{align}
\end{subequations}
wherein~$p$ and~$\vect{v}$ are, respectively, the pressure field and fluid velocity field at position~$\R$ and~$\vect{F}$ is an arbitrary time-dependent force acting tangential to the membrane at position $\vect{r}_0 = h \vect{\hat{z}}$; the unit vector~$\vect{\hat{z}}$ is directed normal to the plane of the undeformed membrane. 
Without loss of generality, we assume throughout this work that $\vect{F} = F \vect{\hat{x}}$.

In an infinitely extended fluid medium, i.e.\ in the absence of the confining elastic membrane, the solution of Eqs.~\eqref{StokesGleischungen} is commonly expressed in terms of the Oseen tensor, also known as the free-space Green's function, or the fundamental solution of Stokes' problem.
Adopting a cylindrical coordinate system~$(r, \phi, z)$, the components of the flow velocity field induced by a transversely oriented Stokeslet acting in an otherwise quiescent fluid medium read~\cite{kim13}
\begin{subequations}
	\begin{align}
		v_r^\mathrm{S} &= \frac{K}{s} \left( 1 + \frac{r^2}{s^2} \right) \cos\phi \, , \\
		v_\phi^\mathrm{S} &= -\frac{K}{s} \, \sin\phi \, , \\
		v_z^\mathrm{S} &= K \, \frac{r \left(z-h\right)}{s^3} \, \cos\phi \, , 
	\end{align}
\end{subequations}
where $s = \left| \R - \R_0 \right| = \left( r^2 + \left(z-h\right)^2 \right)^{1/2}$ denotes the distance from the singularity position and $K=F/ \left(8\pi\eta\right)$ is a dimensional prefactor.
Likewise, the corresponding solution for the pressure field is given by
\begin{equation}
	p^\mathrm{S} = 2 \eta K \, \frac{r}{ s^3} \, \cos\phi \, .
\end{equation}

Owing to the linearity of the Stokes equations~\eqref{StokesGleischungen}, the solution of the hydrodynamic problem for the velocity and pressure fields can be written as a superposition of the solution in an unbounded fluid medium $\{ \vect{v}^\mathrm{S}, p^\mathrm{S} \}$ and a complementary solution (also known as the image system solution) $\{ \vect{v}^*, p^* \}$ that is needed to satisfy the boundary conditions prescribed at the surface of the membrane. 
Accordingly,
\begin{equation}
	\vect{v} = \vect{v}^\mathrm{S} + \vect{v}^* \, , \qquad
	p = p^\mathrm{S} + p^* \, ,    
\end{equation}
where, for an asymmetric flow, the image velocity field~$\vect{v}^*$ can conveniently be expressed as~\cite{shail87}
\begin{equation}
	\vect{v}^* = \bNabla  \Pi + z \bNabla \left( \Psi + \Pi_{,z} \right) 
	- \left( \Psi + \Pi_{,z} \right) \vect{\hat{z}}
	+ \bNabla \times \left( \Omega \, \vect{\hat{z}} \right) \, , \label{vStern}
\end{equation}
with the corresponding image solution for the pressure field
\begin{equation}
	p^* = 2\eta \left( \Psi_{,z} + \Pi_{,zz} \right) \, .
\end{equation}
Commas in indices represent a spatial derivative with respect to
the corresponding coordinate.
Here, $\Pi$, $\Psi$, and~$\Omega$ are harmonic functions satisfying the Laplace equation, i.e., $\boldsymbol{\nabla}^2 \Pi = \boldsymbol{\nabla}^2 \Psi = \boldsymbol{\nabla}^2 \Omega = 0$, the solution of which can generally be expressed in terms of infinite series of Fourier-Bessel integrals~\cite{korenev02}.
By requiring the natural continuity of the fluid velocity at the plane $z=0$ together with the regularity conditions of vanishing velocity and pressure fields as~$\left| \R \right| \to \infty$, the harmonic functions $\Pi$, $\Psi$, and~$\Omega$ can appropriately be expressed for the present geometry in terms of infinite integrals over the wavenumber~$\lambda$ as~\cite{daddi20pof}
\begin{equation}
	\begin{pmatrix}
		\Pi \\ \Psi \\ \Omega
	\end{pmatrix}
	= K
	\begin{pmatrix}
		\cos\phi \\	\cos\phi \\	\sin\phi
	\end{pmatrix}
	\bigintss_0^\infty
	\begin{pmatrix}
		\pi (\lambda) \\ \psi (\lambda) \\ \omega (\lambda)
	\end{pmatrix}
	J_1 \left(\lambda r\right) e^{-\lambda |z|} \, \Intd \lambda \, ,
\end{equation}
with~$J_1$ denoting the first-order Bessel function of the first kind~\cite{abramowitz72}.
In addition, $\pi(\lambda)$, $\psi(\lambda)$, and $\omega(\lambda)$ are unknown wavenumber-dependent functions that will be determined in the following from the boundary conditions prescribed at the plane~$z=0$.

By projecting Eq.~\eqref{vStern} onto the basis vectors of the system of cylindrical coordinates, the radial, azimuthal, and axial components of the image velocity can be expressed in terms of the harmonic functions~$\Pi$, $\Psi$, and~$\Omega$ and their derivatives as
\begin{subequations}\label{ImgVelocities}
	\begin{align}
		v_r^* &= \Pi_{,r} + z \left( \Pi_{,rz} + \Psi_{,r} \right) +  \frac{\Omega_{,\phi}}{r} \, , \label{VelocityR} \\
		v_\phi^* &= \frac{1}{r}  \left( \Pi_{,\phi} + z \left( \Pi_{,\phi z}  + \Psi_{,\phi} \right) \right) - \Omega_{,r} \, , \label{VelocityPhi} \\
		v_z^* &= z \left( \Psi_{,z} + \Pi_{,zz} \right) - \Psi \, . \label{VelocityZ}
	\end{align}
\end{subequations}

\subsection{Elastic membrane model}

The thickness of our finite-sized deformable piece of membrane is assumed to vanish in our description. 
Moreover, the membrane is assumed to be composed of a hyperelastic material that features resistance towards shear and bending.
In-plane shear elasticity of the membrane is modeled by the well-established Skalak model~\cite{skalak73}, which is widely used to represent properties of red blood cell membranes~\cite{krueger11, krueger12, barthes16, baecher17}.
Besides the resistance towards shear deformations, the Skalak model comprises into a single energy functional for the areal strain the resistance of the membrane towards shear and it ensures the local conservation of surface area~\cite{ramanujan98, lac04}.
In addition, bending rigidity is modeled by the celebrated Helfrich model~\cite{helfrich73}, which is represented by a quadratic curvature elastic model~\cite{Guckenberger_jpcm, daddi2017thesis, daddi18epje}.

For a planar elastic membrane, the jumps in linearized traction across the membrane due to shear and bending deformation modes have previously been derived by some of us and provided in Cartesian coordinates in Ref.~\onlinecite{daddi16} as
\begin{align}
	\bigtriangleup \vect{f} = 
	-\frac{\kS}{3} \left( \boldsymbol{\nabla}_\parallel^2 \vect{u} + 
	\left(1+2C\right) \boldsymbol{\nabla}_\parallel
	\varepsilon \right) \notag
	+ \kB \boldsymbol{\nabla}_\parallel^4 u_z \, \vect{\hat{e}}_z \, , 
\end{align}
where~$\kS$ and~$\kB$ denote, respectively, the shear and bending modulus of the membrane, $C$ is the ratio of shear to area dilatation modulus in the Skalak model, and $\vect{u}$ is the displacement vector of the material points composing the membrane relative to their initial positions in the undeformed state.
In addition, $\boldsymbol{\nabla}_\parallel (\cdot) := \vect{\hat{e}}_x \partial (\cdot) / \partial x + \vect{\hat{e}}_y \partial (\cdot) / \partial y$ is the tangential gradient operator and $\varepsilon := \boldsymbol{\nabla}_\parallel \cdot \vect{u}$ is the dilatation function quantifying the relative variation of the surface.

Accordingly, the radial, azimuthal, and axial components of the jumps in traction can, respectively, be expressed in the system of cylindrical coordinates as
\begin{subequations}\label{tractionJumpGleischungen}
	\begin{align}
		\bigtriangleup f_r &= -\frac{\kS}{3} \bigg( 2 \left(1+C\right) \left( u_{r,rr} + \frac{u_{r,r}}{r} - \frac{u_r}{r^2} \right) + \frac{u_{r, \phi\phi}}{r^2} \notag \\
		&\quad-\left(3+2C\right) \frac{u_{\phi, \phi}}{r^2} + \left(1+2C\right) \frac{u_{\phi, r\phi}}{r} 	\bigg) \, , \\
		\bigtriangleup f_\phi &= -\frac{\kS}{3} \bigg( 2\left(1+C\right) \frac{u_{\phi, \phi\phi}}{r^2} + u_{\phi, rr} + \frac{u_{\phi, r}}{r} - \frac{u_\phi}{r^2} \notag \\
		&\quad + \left(3+2C\right) \frac{u_{r,\phi}}{r^2} + \left(1+2C\right) \frac{u_{r,r\phi}}{r} \bigg) \, , \\
		\bigtriangleup f_z &= \kB \bigg( u_{z,rrrr} + \frac{2}{r} \, u_{z, rrr} - \frac{u_{z, rr}}{r^2} + \frac{{u_{z,r}}}{r^3} + \frac{4}{r^4} \, u_{z, \phi\phi} \notag \\
		&\quad- \frac{2}{r^3} \, u_{z, r\phi\phi}
		+\frac{2}{r^2} \, u_{z, rr\phi\phi} + \frac{u_{z, \phi\phi\phi\phi}}{r^4} \bigg) \, .
	\end{align}
\end{subequations}
Notably, the in-plane traction jumps $\bigtriangleup f_r$ and $\bigtriangleup f_\phi$ are solely determined by the shear elasticity of the membrane whereas the out-of-plane traction jump $\bigtriangleup f_z$ is determined by bending resistance only.
It is worth noting that this behavior is in stark contrast to curved membranes where coupling between shear and bending occurs~\cite{daddi17b, daddi17c,daddi17pof, daddi18acta, daddi18creeping, hoell19creeping}.

Having formulated the general solution for the induced flow field and presented a model for the properties of the elastic membrane, we derive in the next section the solution of the overall elastohydrodynamic problem using a formulation in terms of dual integral equations.

\section{Solution of the elastohydrodynamic problem}
\label{sec:loesung}

\subsection{Formulation of the mixed-boundary-value problem}

The boundary conditions imposed at $z=0$ comprise (a)~the discontinuity of the normal stresses at the membrane due to shear and bending resistance modes, and (b)~continuity of the normal stresses in the regions outside the membrane.
They can be presented in a compact form as
\begin{equation}
	\left[ \vect{T}  \right] = \bigtriangleup \vect{f} \, H\left( R-r \right) \, ,  \label{InnerUndOuterProblem} 
\end{equation}
with $H(\cdot)$ denoting the Heaviside step function.
Here, $\vect{T} = \boldsymbol{\sigma} \cdot \vect{\hat{z}}$ stands for the traction vector with $\boldsymbol{\sigma}$ denoting the hydrodynamic viscous stress tensor.
In addition, $[\cdot] = \cdot(z=0^+) - \cdot(z=0^-)$ represents the jump of a given quantity across the membrane.
The components of the stress vector are expressed in cylindrical coordinates in the usual way as
\begin{subequations}\label{StressVector}
	\begin{align}
		T_{r} &= \eta \left( v_{r,z} + v_{z,r} \right) \, , \label{StressRZ} \\
		T_{\phi} &= \eta \left( v_{\phi, z} + \frac{v_{z, \phi}}{r} \right) \, , \label{StressPhiZ} \\
		T_{z} &= -p + 2\eta v_{z,z} \, . \label{StressZZ}
	\end{align}
\end{subequations}

To achieve a closure of the elastohydrodynamic problem at hand, we assume at the surface of the membrane a no-slip boundary condition.
Accordingly, the fluid velocity on the plane $z=0$ is supposed to be equal to the velocity of the material points composing the membrane, i.e., $\dot{\vect{u}} = \vect{v}$ at $z=0$.
In Fourier space, this condition is expressed as $\vect{u} = \vect{v}/ \left(i\omega\right)$, with~$\omega$ denoting the actuation frequency of the system. 
Then, the resulting velocity and pressure jumps across the membrane can be expressed as
\begin{subequations}
	\begin{align}
		[v_{r,z}] &= i\alpha \bigg( 4 \left( v_{r,rr} + \frac{v_{r,r}}{r} - \frac{v_r}{r^2} \right) + B \, \frac{v_{r, \phi\phi}}{r^2}
		-\left(4+B\right) \frac{v_{\phi, \phi}}{r^2} \notag \\
		&\quad \left. + \left(4-B\right) \frac{v_{\phi, r\phi}}{r} 	\bigg) \right|_{z=0} , \\
		[v_{\phi, z}] &= i\alpha \bigg( 4 \, \frac{v_{\phi, \phi\phi}}{r^2} + B \left( v_{\phi, rr} + \frac{v_{\phi, r}}{r} - \frac{v_\phi}{r^2} \right) 
		+ \left(4+B\right) \frac{v_{r,\phi}}{r^2} \notag \\
		&\quad \left. + \left(4-B\right) \frac{v_{r,r\phi}}{r} \bigg) \right|_{z=0} , \\
		\frac{[p]}{\eta} &= 4i \alphaB^3\bigg( v_{z,rrrr} + \frac{2}{r} \, v_{z, rrr} - \frac{v_{z, rr}}{r^2} + \frac{{v_{z,r}}}{r^3} + \frac{4}{r^4} \, v_{z, \phi\phi} \notag \\
		&\quad\left. - \frac{2}{r^3} \, v_{z, r\phi\phi}
		+\frac{2}{r^2} \, v_{z, rr\phi\phi} + \frac{v_{z, \phi\phi\phi\phi}}{r^4} \bigg) \right|_{z=0} ,
	\end{align}
\end{subequations}
where~$B=2/\left(1+C\right)$ is a dimensionless number.
In addition, $\alpha$ and~$\alphaB$ are characteristic length scales associated with shear and bending deformation modes, respectively, defined as~\cite{daddi16}
\begin{equation}
		\alpha  = \frac{\kS}{3B\eta \omega} \, ,  \qquad
		\alphaB = \left( \frac{\kB}{4\eta \omega} \right)^{1/3} \, .
\end{equation}

It is worth noting that, if the membrane undergoes large deformations, the no-slip boundary condition should instead be applied in the displaced state of the membrane, see for instance Refs.~\onlinecite{sekimoto93, weekley06, salez15, saintyves16, rallabandi17prfluids, daddi18stone, rallabandi18, saintyves20, zhang2020direct, kaoui16} where such an approach has been employed.
However, since our attention is restricted here to the membrane behavior in the regime of small displacements characterized by $|\vect{u}| \ll h$, we apply the no-slip condition at $z=0$, confining ourselves to a regime of linearized membrane elasticity.

\subsection{Dual integral equations}

We will show in the sequel that the present mixed-boundary-value problem can be reduced to a system  of dual integral equations for the unknown wavenumber-dependent coefficients~$\pi(\lambda)$, $\psi(\lambda)$, and~$\omega(\lambda)$.
By inserting Eqs.~\eqref{ImgVelocities} into Eqs.~\eqref{InnerUndOuterProblem}, we obtain the following system of integral equations for the inner domain $(0<r<R)$
\begin{subequations}\label{InnerDomainEqs}
	\begin{align}
		\int_0^\infty \lambda^2 \left( \xi_1^+(\lambda) - i\alpha \lambda \xi_2^+(\lambda) \right) J_0 \left(\lambda r\right) \, \Intd \lambda &= g_\mathrm{S}^+ (r) \, , \label{InnerDomainShear1} \\
		\int_0^\infty \lambda^2 \left( \xi_1^-(\lambda) - i\alpha \lambda \xi_2^-(\lambda) \right) J_2 \left(\lambda r\right) \, \Intd \lambda &= g_\mathrm{S}^- (r) \, , \label{InnerDomainShear2} \\
		\int_0^\infty \lambda \psi(\lambda) \left( 1-i\alphaB^3 \lambda^3 \right) J_1 \left(\lambda r\right) \, \Intd \lambda &= g_\mathrm{B} (r) \, , 
		\label{InnerDomainBending}
	\end{align}
\end{subequations}
where we have defined the wavenumber-dependent quantities
\begin{subequations}\label{xi1UNDxi2}
	\begin{align}
		\xi_1^\pm (\lambda) &= 2 \left( 2\pi(\lambda) \pm \omega(\lambda) \right) \, , \label{xi1PMDef} \\
		\xi_2^\pm (\lambda) &= 4\pi(\lambda) \pm B \omega (\lambda) \, , 
	\end{align}
\end{subequations}
in addition to the radial functions
\begin{align}
	g_\mathrm{S}^+ (r) &= \frac{2i\alpha \left( -(B+2)r^4 + \left(B+20\right)h^2r^2 + \left(2B-8\right)h^4 \right)}{\left( r^2 + h^2 \right)^{7/2}} \, , \notag \\
	g_\mathrm{S}^- (r) &= -\frac{6i\alpha r^2 \left( (B-2)r^2 + (B+8)h^2 \right)}{\left(r^2+h^2\right)^{7/2}} \, , \notag \\
	g_\mathrm{B} (r) &= \frac{45i \alphaB^3 hr \left( r^4-12h^2r^2 + 8h^4 \right)}{\left(r^2+h^2\right)^{11/2}} \, . \notag 
\end{align}

Likewise, we obtain the following integral equations for the outer domain $(r>R)$
\begin{subequations}\label{OuterDomainEqs}
	\begin{align}
		\int_0^\infty \lambda^2 \xi_1^+(\lambda) J_0 \left(\lambda r\right) \, \Intd \lambda &= 0 \, , \label{OuterDomainShear1} \\
		\int_0^\infty \lambda^2 \xi_1^-(\lambda) J_2 \left(\lambda r\right) \, \Intd \lambda &= 0 \, , \label{OuterDomainShear2} \\
		\int_0^\infty \lambda \psi(\lambda) J_1 \left(\lambda r\right) \, \Intd \lambda &= 0 \, . \label{OuterDomainBending}
	\end{align}
\end{subequations}

Equations~\eqref{InnerDomainEqs} through \eqref{OuterDomainEqs} form a system of dual integral equations defined on the inner and outer domain boundaries for the unknown wavenumber-dependent functions~$\xi_1^\pm(\lambda)$ and $\psi(\lambda)$.
The latter represent, respectively, those contributions to the image flow field related to shear and bending deformations of the membrane.
Moreover, it follows readily from Eqs.~\eqref{xi1UNDxi2} that 
\begin{equation}
	\xi_2^\pm (\lambda) = \frac{1}{4} 
	\left( \left(2\pm B\right) \xi_1^+(\lambda) + \left(2\mp B\right) \xi_1^-(\lambda) \right) \, . \label{xi2AsFctOfxi1}
\end{equation}

\subsection{Solution for $R\to\infty$}

Before proceeding with solving the system of dual integral equations at hand, we first recover the solution in the limit~$R\to\infty$ corresponding to an infinitely extended planar elastic membrane.
In this situation, the integral equations~\eqref{InnerDomainEqs} are defined in the whole range of values of~$r \in \mathbb{R}$ and can conveniently be solved using inverse Hankel transforms as
\begin{subequations}\label{Hankel}
	\begin{align}
		\lambda \left( \xi_1^+(\lambda) - i\alpha \lambda \xi_2^+(\lambda) \right)
		= \int_0^\infty r g_\mathrm{S}^+(r) J_0(\lambda r) \, \Intd r \, , \\
		\lambda \left( \xi_1^-(\lambda) - i\alpha \lambda \xi_2^-(\lambda) \right)
		= \int_0^\infty r g_\mathrm{S}^+(r) J_2(\lambda r) \, \Intd r \, , \\
		\psi(\lambda) \left( 1-i\alphaB^3 \lambda^3 \right) = \int_0^\infty rg_\mathrm{B}(r) J_1(\lambda r) \, \Intd r \, .
	\end{align}
\end{subequations}
Upon evaluation of the convergent improper integrals forming the right-hand sides of Eqs.~\eqref{Hankel}, we  obtain
\begin{subequations}\label{InvHankelFinal}
	\begin{align}
		\xi_1^\pm (\lambda) - i\alpha \lambda \xi_2^\pm (\lambda)
		&= 2i\alpha \left( 2 \pm B - 2\lambda h \right)e^{-\lambda h} \, , \\
		\psi(\lambda) \left( 1-i\alphaB^3 \lambda^3 \right)
		&= i \alphaB^3 \lambda^4 h e^{-\lambda h} \, .
	\end{align}
\end{subequations}

Finally, by making use of Eqs.~\eqref{xi1UNDxi2} and solving Eqs.~\eqref{InvHankelFinal} for $\pi(\lambda)$, $\omega(\lambda)$, and $\psi(\lambda)$, we find
\begin{subequations}\label{piOmegaPsiInfMem}
	\begin{align}
		\pi(\lambda) &= \frac{i\alpha \left(1-\lambda h\right)}{1-i\alpha \lambda} \, e^{-\lambda h} \, , \\
		\omega(\lambda) &= \frac{2i\alpha B}{2-i\alpha B \lambda} \, e^{-\lambda h} \, , \\
		\psi(\lambda) &= \frac{i\alphaB^3 \lambda^4h}{1-i\alphaB^3 \lambda^3} \, e^{-\lambda h} \, ,
	\end{align}
\end{subequations}
which are in full agreement with the solution for an infinitely extended membrane, as previously derived by some of us~\cite{daddi16, daddi17} using a standard two-dimensional Fourier transform technique~\cite{bickel06, bickel07}.
Notably, in the limits $\alpha\to\infty$ and~$\alphaB\to\infty$, we recover the well-known Blake solution~\cite{blake71} for a transverse Stokeslet acting near a planar hard wall.

Thanks to the decoupled nature of shear and bending effects, the solution of the elastohydrodynamic problem can adequately be obtained by considering these deformation modes independently.
It is worth noting that a coupling behavior occurs in elastic membranes with finite curvature~\cite{daddi17b, daddi17c}, or for two parallel~\cite{daddi16b}, thermally warped~\cite{Kosmrlj14}, or closely coupled~\cite{auth07} fluctuating membranes.

\subsection{Shear contribution}

Following the recipes by Copson~\cite{copson61}, we express the solution of the resulting system of dual integral equations stated by Eqs.~\eqref{InnerDomainEqs} and \eqref{OuterDomainEqs} for the sought-for wavenumber-dependent functions~$\xi_1^\pm (\lambda)$ as
\begin{subequations}\label{xi1SolutionForm}
	\begin{align}
		\xi_1^+ (\lambda) &= \lambda^{-\frac{3}{2}} \int_0^R
				\hat{\xi}_1^+ (t) J_{\frac{1}{2}} \left( \lambda t \right) \, \Intd t \, , \\
		\xi_1^- (\lambda) &= \lambda^{-\frac{3}{2}} \int_0^R
				\hat{\xi}_1^- (t) J_{\frac{5}{2}} \left( \lambda t \right) \, \Intd t \, , 
	\end{align}
\end{subequations} 
where~$\hat{\xi}_1^\pm (t), t \in [0,R],$ are unknown functions that need to be determined.
It can be checked that the integral equations for the outer problem given by Eqs.~\eqref{OuterDomainShear1} and \eqref{OuterDomainShear2} are satisfied by making use of the classic identity by Watson~\cite{watson95},
\begin{equation}
	\int_0^\infty \lambda^{s} J_p(a\lambda) J_q(b\lambda) \Intd \lambda =
	\frac{b^q H(a-b)}{a^p \Gamma \left(p-q\right)} \left(\frac{a^2-b^2}{2}\right)^{-s} ,
	\label{watson} 
\end{equation}
where $s=1+q-p < 1$ and $\Gamma(\cdot)$ denotes Euler's Gamma function~\cite{abramowitz72}.
The integral representations of~$\xi_2^\pm (\lambda)$ are then obtained by inserting Eqs.~\eqref{xi1SolutionForm} into Eqs.~\eqref{xi2AsFctOfxi1}.

Next, by inserting Eqs.~\eqref{xi1SolutionForm} into the system of dual integral equations for the inner problem stated by Eqs.~\eqref{InnerDomainShear1} and~\eqref{InnerDomainShear2}, and interchanging the order of the integrations with respect to the variables $\lambda \in [0,\infty)$ and~$t \in [0,R]$, we obtain
\begin{widetext}
\begin{subequations}\label{SysDualIntStart}
	\begin{align}
		\int_0^R Q_\halb^0 (r,t) \, \hat{\xi}_1^+(t) \, \Intd t 
		-i\alpha \left( 
		b^+ \int_0^R K_\halb^0 (r,t) \, \hat{\xi}_1^+ (t) \, \Intd t
		+ b^- \int_0^R K_\fivehalb^0 (r,t) \, \hat{\xi}_1^- (t) \, \Intd t
		\right) &= g_\mathrm{S}^+ (r) \, , \label{InnerTransShear1} \\
		\int_0^R Q_\fivehalb^2 (r,t) \, \hat{\xi}_1^-(t) \, \Intd t 
		-i\alpha \left( 
		b^- \int_0^R K_\halb^2 (r,t) \, \hat{\xi}_1^+ (t) \, \Intd t
		+ b^+ \int_0^R K_\fivehalb^2 (r,t) \, \hat{\xi}_1^- (t) \, \Intd t
		\right) &= g_\mathrm{S}^- (r) \, , \label{InnerTransShear2}
	\end{align}
\end{subequations}
\end{widetext}
where we have defined $b^\pm = \left(2 \pm B\right)/4$ in addition to the kernel functions
\begin{subequations}
	\begin{align}
		Q_p^q (r,t) &= \int_0^\infty \lambda^\halb J_p(\lambda t) J_q(\lambda r) \, \Intd \lambda \, , \label{QpqDef} \\
		K_p^q (r,t) &= \int_0^\infty \lambda^\threehalb J_p(\lambda t) J_q(\lambda r) \, \Intd \lambda \, .
	\end{align}
\end{subequations}

An analytical treatment of the integral terms involving~$Q_p^q$ is straightforward.
To deal with the terms involving~$K_p^q$, we use integration by parts to express them in terms of integrals involving kernels of the type~$Q_p^q$, which we know well how to handle.
By making use of the recurrence relations~\cite{abramowitz72}
\begin{subequations}
	\begin{align}
		J_{\alpha-1} (x) &= x^{-\alpha} \frac{\Intd}{\Intd x} \left( x^\alpha J_\alpha (x) \right) \, , \\
		J_{\alpha+1} (x) &= -x^\alpha \frac{\Intd}{\Intd x} \left( x^{-\alpha} J_\alpha (x) \right) \, , 
	\end{align}
\end{subequations}
it follows that
\begin{subequations}
	\begin{align}
		\int_0^R & \hat{\xi}_1^+(t) \, J_\halb (\lambda t) \, \Intd t \notag \\
		&= \lambda^{-1} \int_0^R \hat{\xi}_1^+(t) \, t^{-\threehalb} 
		\frac{\Intd}{\Intd t} \left( t^\threehalb J_\threehalb (\lambda t) \right) \Intd t \notag \\
		&= \lambda^{-1} \left( \hat{\xi}_1^+(R) J_\threehalb(\lambda R) 
		- \int_0^R \hat{X}^+(t) J_\threehalb (\lambda t) \, \Intd t \right) , 
		\label{IntegrationByPartsBesselHalb}
		\\
		\int_0^R & \hat{\xi}_1^-(t) \, J_\fivehalb (\lambda t) \, \Intd t \notag \\
		&= -\lambda^{-1} \int_0^R \hat{\xi}_1^-(t) \, t^{\threehalb} 
		\frac{\Intd}{\Intd t} \left( t^{-\threehalb} J_\threehalb (\lambda t) \right) \Intd t \notag \\
		&= \lambda^{-1} \left( \int_0^R \hat{X}^-(t) J_\threehalb (\lambda t) \, \Intd t - \hat{\xi}_1^-(R) J_\threehalb(\lambda R) \right) ,
	\end{align}
\end{subequations}
where we have defined
\begin{subequations}\label{XP_XM}
	\begin{align}
		\hat{X}^+(t) &= t^\threehalb \frac{\Intd}{\Intd t} 
		\left( t^{-\threehalb} \hat{\xi}_1^+(t) \right) \, , \\
		\hat{X}^-(t) &= t^{-\threehalb} \frac{\Intd}{\Intd t} 
		\left( t^{\threehalb} \hat{\xi}_1^-(t) \right) \, , 
	\end{align}
\end{subequations}
and where we have assumed that $t^\threehalb \hat{\xi}_1^\pm(t) \to 0$ as $t \to 0$. 
Equations~\eqref{XP_XM} can be presented in a compact and simplified form as
\begin{equation}
	\hat{X}^\pm (t) = \frac{\Intd}{\Intd t} \, \hat{\xi}_1^\pm (t)
	\mp \frac{3}{2t} \, \hat{\xi}_1^\pm (t) \, .
\end{equation}

Combining results, the system of dual integral equations stated by Eqs.~\eqref{SysDualIntStart} can be expressed in terms of integrals involving kernels of the type~$Q_p^q$ as
\begin{widetext}
\begin{subequations}
	\begin{align}
			\int_0^R Q_\halb^0 (r,t) \, \hat{\xi}_1^+(t) \, \Intd t -i\alpha 
			\left( 
			\left( b^+ \hat{\xi}_1^+(R) - b^- \hat{\xi}_1^-(R) \right) Q_\threehalb^0(r,R)
			- \int_0^R \left( b^+ \hat{X}^+(t) - b^- \hat{X}^-(t) \right)
			Q_\threehalb^0 (r,t) \, \Intd t	\right) = g_\mathrm{S}^+ (r) \, , \\
			\int_0^R Q_\fivehalb^2 (r,t) \, \hat{\xi}_1^-(t) \, \Intd t -i\alpha 
			\left( 
			\left( b^- \hat{\xi}_1^+(R) - b^+ \hat{\xi}_1^-(R) \right) Q_\threehalb^2(r,R)
			- \int_0^R \left( b^- \hat{X}^+(t) - b^+ \hat{X}^-(t) \right)
			Q_\threehalb^2 (r,t) \, \Intd t	\right) = g_\mathrm{S}^- (r) \, .
		\end{align}
\end{subequations}
The latter system of equations can be presented in the final simplified form
\begin{subequations}\label{SysDualIntFinal}
	\begin{align}
		\int_r^R t^{-\halb} \left( t^2-r^2 \right)^{-\halb} \hat{\xi}_1^+ (t) \, \Intd t
		-i\alpha \left( 
		 \mathcal{K}
		-\int_0^r \varphi(r, t) \, \hat{Y}^+(t) \, \Intd t 
		-\tfrac{\pi}{2} \int_r^R t^{-\threehalb} \hat{Y}^+(t) \, \Intd t \right)
		&= \left( \tfrac{\pi}{2} \right)^\halb g_\mathrm{S}^+(r) \, , \\
		r^2 \int_r^R t^{-\fivehalb} \left( t^2-r^2 \right)^{-\halb} \hat{\xi}_1^- (t) \, \Intd t
		+ i\alpha r^{-2} \int_0^r t^\threehalb \left( r^2-t^2 \right)^{-\halb} \hat{Y}^-(t)  \, \Intd t &= \left( \tfrac{\pi}{2} \right)^\halb g_\mathrm{S}^-(r) \, , 
	\end{align}	
\end{subequations}
\end{widetext}
where we have defined the constant
\begin{equation}
	\mathcal{K} = \tfrac{\pi}{2} \, R^{-\threehalb} \left( b^+ \hat{\xi}_1^+(R) - b^- \hat{\xi}_1^-(R) \right) ,
\end{equation}
in addition to the abbreviations
\begin{subequations}
	\begin{align}
		\hat{Y}^\pm (t) &= b^\pm \hat{X}^+(t) - b^\mp \hat{X}^-(t) \, , \\
		\varphi(r,t) &= t^{-\halb} \left( t^{-1} \arcsin \left( \tfrac{t}{r} \right) - \left(r^2-t^2\right) ^{-\halb}\right) .
	\end{align}
\end{subequations}

Here, we have made use of Watson's identity stated by Eq.~\eqref{watson}, to obtain
\begin{subequations}
	\begin{align}
		Q_\halb^0 (r,t) &= 
		\left( \tfrac{2}{\pi t} \right)^\halb \left( t^2-r^2 \right)^{-\halb} H(t-r)\, , \\
		Q_\fivehalb^2 (r,t) &= 
		\left( \tfrac{2}{\pi t} \right)^\halb
		\left(\tfrac{r}{t}\right)^2 \left(t^2-r^2\right)^{-\halb} H(t-r)\, ,
	\end{align}
\end{subequations}
together with
\begin{subequations}
	\begin{align}
		Q_\threehalb^0 (r,t) &= 
			\left( \tfrac{2}{\pi t} \right)^\halb 
			\left( t^{-1} \arcsin \left( \tfrac{t}{r} \right) - \left(r^2-t^2\right) ^{-\halb}\right) H (r-t) \notag \\
			&\quad+ \left( \tfrac{\pi}{2t} \right)^\halb t^{-1} H(t-r) \, , \\
		Q_\threehalb^2 (r,t) &=
		\left( \tfrac{2}{\pi t} \right)^\halb
		\left(\tfrac{t}{r}\right)^2 \left( r^2-t^2 \right)^{-\halb} 
		H (r-t) \, . \label{Q_threehalb_2}
	\end{align}
\end{subequations}
It is worth noting that Eq.~\eqref{Q_threehalb_2} implies $Q_\threehalb^2(r,R) = 0$.

Due to the somehow complicated nature of the resulting system of integro-differential equations given by Eqs.~\eqref{SysDualIntFinal}, an analytical solution is far from being trivial.
Therefore, recourse to numerical methods is necessary and indispensable. 
However, we will show in the sequel that for an elastic membrane with infinite resistance toward shear, an analytical solution is fortunately straightforward. 
For that aim, we use the standard series expansion solution technique~\cite{daddi20pof}.
Accordingly, we expand the known radial functions on the right-hand sides of Eqs.~\eqref{SysDualIntFinal} as Taylor series about the origin and express the solutions of the system of dual integral equations as infinite power series with unknown coefficients of the form
\begin{equation}\label{xi1Expansion}
	\hat{\xi}_1^\pm (t) = \left( \tfrac{t}{2\pi} \right)^\halb
	\sum_{n \ge 0} c_n^\pm \left( \tfrac{t}{h} \right)^{2n+1} \, .
\end{equation}
By solving for the unknown series coefficients by identification of terms of the same power of~$r$, we readily obtain
\begin{subequations}\label{cn}
	\begin{align}
		c_n^+ &= 16(n+1)(2n+1) \qquad \text{for } n \ge 1 \, , \\
		c_n^- &= 32n (n+1) \, , 
	\end{align}
\end{subequations}
and
\begin{align}\label{c0}
	c_0^+ &= \frac{48}{h^2} \frac{2-B}{2+B} \sum_{n \ge 1} (-1)^n (n+1) \left( \frac{R}{h} \right)^{2n} \notag \\
		  &= -\frac{48(2-B) \left( R^2+2h^2 \right)}{(2+B) \left( R^2+h^2 \right)^2} \left( \frac{R}{h} \right)^2 \, .
\end{align}

Finally, by inserting the expressions of the series coefficients given by Eqs.~\eqref{cn} and \eqref{c0} into Eq.~\eqref{xi1Expansion}, and evaluating the infinite sums analytically, we obtain
\begin{subequations}
	\begin{align}
		\hat{\xi}_1^+ (t) &= 16\left( \frac{t}{2\pi} \right)^\halb
		\left( \delta t - \frac{t^3 \left( t^4+3h^2 t^2 + 6h^4 \right)}{h^3 \left( t^2 + h^2 \right)^3} \right) , \\
		\hat{\xi}_1^- (t) &= 64 \left( \frac{t}{2\pi} \right)^\halb
		\frac{h t^3}{\left( t^2 + h^2 \right)^3} \, , 
	\end{align}
\end{subequations}
where
\begin{equation}
	\delta = \frac{1}{(2+B)h^3} \left( 2\left(4-B\right) - \frac{3(2-B)R^2 \left(R^2+2h^2\right)}{\left( R^2 + h^2 \right)^2} \right) .
\end{equation}

In particular, it can be checked that the solution near an infinitely extended membrane that only allows for shear deformations is recovered in the limit $R\to\infty$.

\subsection{Bending contribution}

We proceed in an analogous way as for the shear-related contribution to the image flow field and express the solution of the dual integral equations for the bending-related wavenumber-dependent function~$\psi(\lambda)$ as
\begin{equation} \label{psiSolutionForm}
	\psi(\lambda) = \lambda^{-\halb} \int_0^R \hat{\psi}(t) J_\threehalb (\lambda t) \, \Intd t \, , 
\end{equation}
where $\hat{\psi}(t),~t \in [0,R],$ is an unknown function that needs to be determined.
Equation~\eqref{psiSolutionForm} clearly satisfies Eq.~\eqref{OuterDomainBending} for the outer problem upon making use of Watson's identity stated by Eq.~\eqref{watson}.
Inserting this form into Eq.~\eqref{InnerDomainBending} for the inner problem yields 
\begin{equation}
	\int_0^R \hat{\psi}(t) \, \Intd t 
	\int_0^\infty \lambda^\halb \left( 1-i\alphaB^3 \lambda^3 \right) J_\threehalb(\lambda t)
	J_1(\lambda r) \, \Intd \lambda = g_\mathrm{B}(r) \, .
	\label{BendingIntEqnStart}
\end{equation}
The evaluation of the first term on the left-hand side of Eq.~\eqref{BendingIntEqnStart} can readily be taken care of.
Nevertheless, the evaluation of the second term is more challenging. 
This can conveniently be dealt with by performing three successive integrations by parts to reduce the degree of the prefactor~$\lambda^3$ recursively by one degree at a time so as to bring the integral equation into a more familiar form, amenable to further analytical treatment. 
A first integration by parts yields
\begin{align}
	\int_0^R & \hat{\psi}(t) J_\threehalb(\lambda t) \, \Intd t \notag \\
	&= -\lambda^{-1} \int_0^R \hat{\psi}(t) \, 
	t^\halb \frac{\Intd}{\Intd t} \left( t^{-\halb} J_\halb (\lambda t) \right) \Intd t \notag \\
	&= \lambda^{-1} \left( \int_0^R \hat{P}(t) \, J_\halb (\lambda t) \, \Intd t
	-\hat{\psi} (R) \, J_\halb (\lambda R) \right) , \label{IntegrationByPartsBesselThreehalb}
\end{align}
where we have assumed that $t^\halb \hat{\psi} (t) \to 0$ as $t\to 0$ and defined 
\begin{equation}
	\hat{P}(t) = t^{-\halb} \frac{\Intd}{\Intd t} 
	\left( t^\halb \hat{\psi}(t) \right) =
	\hat{\psi}'(t) + \frac{1}{2t} \, \hat{\psi}(t) \, ,
\end{equation}
with prime denoting a derivative with respect to the argument. 
Next, following the same procedure as in Eq.~\eqref{IntegrationByPartsBesselHalb}, a second integration by parts gives
\begin{align}
	\int_0^R & \hat{P}(t) \, J_\halb (\lambda t) \, \Intd t \notag \\
	&= \lambda^{-1} \left( \hat{P}(R) \, J_\threehalb(\lambda R)
	- \int_0^R \hat{Q}(t) \, J_\threehalb(\lambda t) \, \Intd t	\right),
\end{align}
where we have assumed that $t^\threehalb \hat{P} (t) \to 0$ as $t \to 0$ and defined
\begin{equation}
	\hat{Q}(t) = t^\threehalb \frac{\Intd}{\Intd t} 
	\left( t^{-\threehalb} \hat{P} (t) \right) 
	= \hat{\psi}''(t) - \frac{\hat{\psi}'(t)}{t} - \frac{5}{4t^2} \, \hat{\psi}(t) \, .
\end{equation}
Finally, following the same procedure as in Eq.~\eqref{IntegrationByPartsBesselThreehalb}, a third integration by parts leads to
\begin{align}
	\int_0^R & \hat{Q}(t) \, J_\threehalb(\lambda t) \, \Intd t \notag \\
	&=\lambda^{-1} \left( 
	\int_0^R \hat{S}(t) \, J_\halb(\lambda t) \, \Intd t
	-\hat{Q}(R) \, J_\halb (\lambda R) \right) ,
\end{align}
where we have assumed that $t^\halb \hat{Q}(t) \to 0$ as $t \to 0$ and defined 
\begin{align}
	\hat{S}(t) &= t^{-\halb} \frac{\Intd}{\Intd t} 
	\left( t^\halb \hat{Q}(t) \right) \notag \\
	&=\hat{\psi}'''(t) - \frac{1}{2t} \, \hat{\psi}''(t) - \frac{3}{4t^2} \, \hat{\psi}'(t) + \frac{15}{8t^3} \, \hat{\psi} (t) \, .
	\label{Shat}
\end{align}

To ensure convergence of the underlying infinite integrals, we require that $\hat{\psi} (R) = 0$ and $\hat{P} (R) = 0$ so that $\psi'(R) = 0$.
Then, by collecting results, Eq.~\eqref{BendingIntEqnStart} can be cast in the form
\begin{equation}
	\int_0^R Q_\threehalb^1 (r,t) \, \hat{\psi}(t) \, \Intd t 
	+ i\alphaB^3 \int_0^R Q_\halb^1(r,t) \, \hat{S}(t) \, \Intd t = g_\mathrm{B}(r) ,
	\label{IntBenAlmostFinal}
\end{equation}
wherein the kernel functions~$Q_p^q (r,t)$ have been defined above by Eqs.~\eqref{QpqDef}.
It follows from Watson's identity stated by Eq.~\eqref{watson} that
\begin{subequations}
	\begin{align}
		Q_\threehalb^1 (r,t) &= \left( \tfrac{2}{\pi t} \right)^\halb
		r t^{-1} \left( t^2 - r^2 \right)^{-\halb} H (t-r) \, , \\
		Q_\halb^1 (r,t) &= \left( \tfrac{2t}{\pi} \right)^\halb r^{-1} \left( r^2-t^2 \right)^{-\halb} H(r-t) \, .
	\end{align}
\end{subequations}

Combining results, Eq.~\eqref{IntBenAlmostFinal} can be presented in the final simplified form
\begin{align}
	r \int_r^R & t^{-\threehalb} \left( t^2 - r^2 \right)^{-\halb} \hat{\psi} (t) \, \Intd t + i \alphaB^3 \notag \\
	&\times r^{-1} \int_0^r t^\halb  \left( r^2-t^2 \right)^{-\halb} \hat{S}(t) \, \Intd t = \left( \tfrac{\pi}{2} \right)^\halb  g_\mathrm{B} (r) \, .
	\label{BendingIntEqnFinal}
\end{align}

Equation~\eqref{BendingIntEqnFinal} is an integro-differential equation for the unknown function~$\hat{\psi}(t)$.
Since the complexity of the kernel functions precludes an analytical solution, the integral equations will thus be solved numerically.
In particular, for a stiff membrane with infinite resistance toward bending, Eq.~\eqref{BendingIntEqnFinal} can be reduced into a classic Volterra integral equation of the first kind~\cite{carleman21, smithies58, anderssen80, arfken95}, the solution of which can be obtained analytically using standard approaches.

It turns out that the resulting integral equation for infinite shear and bending rigidities can be mapped into the Abel integral equation.
We recall for the sake of completeness that the Abel integral equation is stated by
\begin{equation}
	\int_0^r g(t) \left( r^2 - t^2 \right)^{-\halb} \, \Intd t = f(r) \, , \label{Abel}
\end{equation}
wherein $g(r)$ and $f(r)$ are the unknown and known functions, respectively.
If $f(r)$ is continuously differentiable at any value of~$r$, then Eq.~\eqref{Abel} has a unique continuous solution given by~\cite{whittaker96, carleman22, tamarkin30}
\begin{equation}
	g(r) = \frac{2}{\pi} \frac{\Intd}{\Intd r} \int_0^r f(t) \left( r^2-t^2 \right)^{-\halb} t \, \Intd t \, .
\end{equation}
Accordingly, in the limit of $\alphaB \to \infty$, the solution of Eq.~\eqref{BendingIntEqnStart} is given by
\begin{equation}
	\hat{S}(t) = 240 \left( \frac{2t}{\pi} \right)^\halb
	\frac{h^2 t \left(3t^2-h^2\right) \left(t^2-3h^2\right)}{\left( t^2+h^2 \right)^6} \, . \label{DiffEqShat}
\end{equation}
We recall that~$\hat{S}(t)$ has been expressed above as a function of~$\hat{\psi} (t)$ and its highest derivatives by Eq.~\eqref{Shat}.
Thus, Eq.~\eqref{DiffEqShat} represents a third-order ordinary differential equation for the unknown function~$\hat{\psi}(t)$ subject to the boundary conditions $t^\halb \hat{\psi}(t) \to 0$ and $t^\threehalb \hat{\psi}'(t) \to 0$ as $t\to 0$ in addition to $\hat{\psi}(R) = 0$, the solution of which is obtained as
\begin{equation}
	\hat{\psi}(t) = \left( \frac{2t}{\pi} \right)^\halb
	\frac{8h^2t^2 \left(t^2-R^2\right) W(t)}{\left( t^2+h^2 \right)^3 \left( R^2+h^2 \right)^3} \, , 
\end{equation}
where
\begin{equation}
	W(t) = t^4 + \left(R^2+3h^2\right)t^2 + R^4 + 3R^2h^2 + 3h^4 \, .
\end{equation}

It is worth mentioning that the limit of~$\psi'(t)$ as $t$ tends to~$R$ from the right does not necessarily vanish for all values of~$h$ and~$R$.
Therefore, $\psi$ is a continuously differentiable function only in the semi-open interval $[0,R)$.
This behavior is due to the fact that Helfrich's model is only defined in principle for closed topological manifolds so that boundary effects have not been properly included in the bending model used in this work.
It can readily be verified that the solution near an infinitely extended membrane with pure bending resistance is recovered in the limit $R\to\infty$.

As a further remark, we briefly address the limit of vanishing frequency
$\omega$. Our deformable piece of membrane is free to move and not
anchored. At the same time, we assume it to be located around $z=0$.
Because of the nature of low-Reynolds-number flows, the membrane for
$\omega\neq0$ will oscillate around its initial position, where $| \vect{F}  |$
needs to be chosen in a way to keep these oscillations small enough in
magnitude for our analysis to remain quantitative. Still, strictly
speaking, a steady state $\omega=0$ does not exist, as the membrane will
perform growing net translations and rotations. However, the limit
$\omega\rightarrow0$ is smooth, and in this sense we can perform a
corresponding analysis for small enough~$| \vect{F}  |$.

Having formulated the solution of the asymmetric flow problem in terms of integro-differential equations and provided analytical solutions for infinite shear and bending rigidities, we will use in the next section the derived solution to assess the effect of the confining elastic membrane on the translational motion of a nearby point-like particle located in its vicinity.

\section{Hydrodynamic mobility function}
\label{sec:mobilitaet}

The exact calculation of the flow field presented in the previous section can be employed to probe the effect of the membrane on the hydrodynamic drag acting on a nearby spherical particle.
This effect is commonly quantified by the hydrodynamic mobility function which relates the velocity of a particle to the hydrodynamic force exerted on its surface~\cite{swan07, swan10}.
In a bulk Newtonian fluid of constant dynamic viscosity~$\eta$, the translational mobility of a spherical particle of radius~$a$ is given by the familiar Stokes law~\cite{stokes51} as~$\mu_0 = 1/(6\pi\eta a)$.
The leading-order correction to the mobility is obtained by evaluating the image flow field at the particle position as
\begin{equation}
	\Delta\mu = F^{-1}\lim_{(r,z) \to (0, h)} v_\parallel^* \, , 
	\label{DeltaMuDef}
\end{equation}
with the translational velocity parallel to the membrane 
\begin{equation}
	v_\parallel^* \equiv v_x^* = v_r^* \cos\phi - v_\phi^* \sin\phi \, . \label{VelocityParallel}
\end{equation}
Scaling by the bulk mobility~$\mu_0$, the scaled correction to the hydrodynamic mobility can be written as
\begin{equation}
	\frac{\Delta\mu}{\mu_0} = - k \, \frac{a}{h} \, . \label{CorrectionfaktorDef}
\end{equation}
Here, $k = k_\mathrm{S} + k_\mathrm{B}$ is a dimensionless number known as the scaled correction factor to the Stokes steady drag, where $k_\mathrm{S}$ and $k_\mathrm{B}$ are contributions stemming from shear and bending deformation modes of the membrane, respectively.
By inserting Eqs.~\eqref{VelocityR} and~\eqref{VelocityPhi} into \eqref{DeltaMuDef}, the shear- and bending-related parts of the correction factors can be written in an integral form as
\begin{subequations}
	\begin{align}
		k_\mathrm{S} &= \frac{3}{8}
		\int_0^\infty \lambda h \left( \left(\lambda h - 1\right)\pi(\lambda) - \omega(\lambda) \right) e^{-\lambda h} \, \Intd \lambda \, , \\
		k_\mathrm{B} &= \frac{3}{8}
		\int_0^\infty -\lambda h^2 \psi(\lambda) \, e^{-\lambda h} \, \Intd \lambda \, .
	\end{align}
\end{subequations}
It follows from Eqs.~\eqref{xi1PMDef} that 
\begin{subequations}\label{piUNDomega}
	\begin{align}
		\pi(\lambda) &= \frac{1}{8} \left( \xi_1^+(\lambda) + \xi_1^-(\lambda) \right) , \\
		\omega(\lambda) &= \frac{1}{4} \left( \xi_1^+(\lambda) - \xi_1^-(\lambda) \right) .
	\end{align}
\end{subequations}
Here, $\xi_1^\pm(\lambda)$ depend on the characteristic length scales associated with shear~$\alpha = \kS / \left(3B\eta\omega\right)$.
They are obtained by solving numerically the system of integro-differential equations~\eqref{SysDualIntFinal} and using the integral representations stated by Eqs.~\eqref{xi1SolutionForm}.
In addition, $\psi(\lambda)$ depends on the characteristic length scale associated with bending~$\alphaB = \left( \kB / \left(4\eta\omega\right) \right)^\frac{1}{3}$ and can be obtained by solving numerically the integro-differential equations~\eqref{BendingIntEqnFinal} and using the integral representation given by Eq.~\eqref{psiSolutionForm}.
Accordingly, the correction factors $k_\mathrm{S}$ and $k_\mathrm{B}$ cannot be expressed analytically in terms of~$\alpha$ and~$\alphaB$.
In the following, we chose to discuss the variations of $k_\mathrm{S}$ and $k_\mathrm{B}$ as functions of the scaled frequencies associated with shear and bending, $\beta$ and~$\betaB$, respectively defined as
\begin{align}
	\beta = \frac{2h}{\alpha} = \frac{6Bh\eta\omega}{\kS} \, , \qquad
	\betaB^3 = \left(\frac{2h}{\alphaB}\right)^3 = \frac{32h^3\eta\omega}{\kB} \, .
\end{align}
In this way, the dependence of the mobility correction on the shear and bending moduli is incorporated by holding the frequency fixed and varying the moduli.

Then, the part of the scaled correction factor associated with shear and bending of the membrane can be expressed in the final form as
\begin{subequations}\label{kSkB}
	\begin{align}
		k_\mathrm{S} &= \frac{3h^\halb}{64} \int_0^R 
		\bigg( \left( Z_\halb^+(\tau)  - 3Z_\halb^-(\tau) \right) \hat{\xi}_1^+(t) \notag \\
		&\quad+ \left( Z_\fivehalb^+(\tau) + Z_\fivehalb^-(\tau) \right) \hat{\xi}_1^-(t) \bigg) \Intd t \, , \label{kS} \\
		k_\mathrm{B} &= -\frac{3h^\halb}{8} \int_0^R Z_\threehalb^+(\tau) \, \hat{\psi}(t) \, \Intd t \, , \label{kB}
	\end{align}
\end{subequations}
wherein $\tau = t/h$ is a scaled variable.
Moreover, 
\begin{equation}
	Z_p^\pm (\tau) = \int_0^\infty u^{\pm \halb} e^{-u} J_p(\tau u) \, \Intd u \, .
\end{equation}
From here, we calculate
\begin{subequations}
	\begin{align}
		Z_\halb^- (\tau) &= \left( \tfrac{2}{\pi \tau} \right)^\halb
		\arctan \tau \, , \\
		Z_\halb^+ (\tau) &= \left( \tfrac{2\tau}{\pi} \right)^\halb 
		\left( 1+\tau^2 \right)^{-1} \, , \\
		Z_\fivehalb^-(\tau) &= \left( 2\pi \tau \right)^{-\halb}
		\left( \left( 1+3\tau^{-2} \right) \arctan\tau - 3 \tau^{-1} \right) , \\
		Z_\fivehalb^+(\tau) &= \left( \tfrac{2}{\pi} \right)^\halb
		\tau^{-\threehalb} \left( \frac{3+2\tau^2}{1+\tau^2} - 3\tau^{-1} \arctan\tau \right), \\
		Z_\threehalb^+(\tau) &= \left( \tfrac{2}{\pi} \right)^\halb \tau^{-\threehalb}
		\left( \arctan\tau - \frac{\tau}{1+\tau^2} \right).
	\end{align}
\end{subequations}

In particular, for infinite shear and bending rigidities, Eqs.~\eqref{kSkB} can be integrated analytically to yield closed analytical expressions of the correction factors, namely
\begin{subequations}\label{DeltaMuFinal}
	\begin{align}
		k_\mathrm{S}^\infty &= \frac{\epsilon \left( 1+\epsilon^2\right)^{-1} f_1 + 3f_2 \arctan \left( \epsilon^{-1} \right)}{16\pi (2+B) \left( 1+\epsilon^2 \right)^2} \, , \label{DeltaMuShearFinal} \\
		k_\mathrm{B}^\infty &= \frac{\epsilon g_1 + 3g_2\arctan \left( \epsilon^{-1} \right)}{16\pi \left(1+\epsilon^2\right)^3} \, ,  \label{DeltaMuBendingFinal}
	\end{align}
\end{subequations}
where we have defined the dimensionless number
\begin{equation}
	\epsilon = \frac{h}{R} \, , 
\end{equation}
in addition to
\begin{subequations}
	\begin{align}
		f_1 &= 15 \left( 2+B \right) + 4\left(56-17B\right)\epsilon^2 + 9\left(18-11B\right)\epsilon^4 \, , \\
		f_2 &= 5\left( 2+B \right) - 6\left(10-B\right)\epsilon^2 - \left( 54-33B \right) \epsilon^4 
	\end{align}
\end{subequations}
for the shear-related part and 
\begin{subequations}
	\begin{align}
		g_1 &= 165\epsilon^4 + 8\epsilon^2 + 3 \, , \\
		g_2 &= \left(1+5\epsilon^2\right) \left( 1-2\epsilon^2 - 11\epsilon^4 \right)
	\end{align}
\end{subequations}
for the bending-related part.

For a significantly large membrane such that $\epsilon \ll 1$, Eqs.~\eqref{DeltaMuFinal} can be expanded in power series of~$\epsilon$ as
\begin{subequations}\label{seriesKInf}
	\begin{align}
		k_\mathrm{S}^\infty &= \frac{15}{32} - \frac{3}{8}\frac{20+B}{2+B} \, \epsilon^2  
		 + \frac{6}{\pi} \frac{4-B}{2+B} \, \epsilon^3
		+ \mathcal{O} \left( \epsilon^4 \right) \, , \\
		k_\mathrm{B}^\infty &= \frac{3}{32} - \frac{9}{4} \, \epsilon^4 + \frac{72}{5\pi} \, \epsilon^5
		+ \mathcal{O} \left( \epsilon^6 \right) .
	\end{align}
\end{subequations}
Notably, we recover in the limit $\epsilon\to 0$ the correction factors $15/32$ and~$3/32$ near an infinitely extended planar elastic membrane endowed with shear and bending rigidities, respectively~\cite{daddi16}. 
By summing up both contributions, we further recover in the limit $\epsilon\to 0$ the familiar leading-order correction factor $9/16$ near a plane solid wall bounding a semi-infinite, otherwise quiescent fluid, as originally obtained by Lorentz using the reciprocal theorem~\cite{lorentz07,happel12,cichocki98}.

Besides, the hydrodynamic mobility for asymmetric motion tangent to a stationary no-slip disk has previously been obtained by Miyazaki~\cite{miyazaki84}.
The latter made use of the Green and Neumann functions supplemented by the edge function technique to obtain closed form expressions for the solution of the creeping flow induced by a Stokeslet acting close to a circular no-slip disk.
Using our notation, the scaled correction factor is expressed by 
\begin{equation}
	k_\text{Disk} = \frac{9}{16} + \frac{1}{8\pi} \left( \frac{\epsilon\left(9-\epsilon^2\right)}{\left(1+\epsilon^2\right)^2} - 9\arctan \epsilon \right) . \label{DeltaMuHard}
\end{equation}
For a very large hard disk of $\epsilon \ll 1$, expanding in Taylor series about~$\epsilon$ yields
\begin{equation}
	k_\text{Disk}  = \frac{9}{16} - \frac{2}{\pi} \, \epsilon^3 + \frac{17}{5\pi} \, \epsilon^5 + \bigO \left( \epsilon^7 \right) . \label{seriesKHard}
\end{equation}

\begin{figure}
	\includegraphics[scale=1]{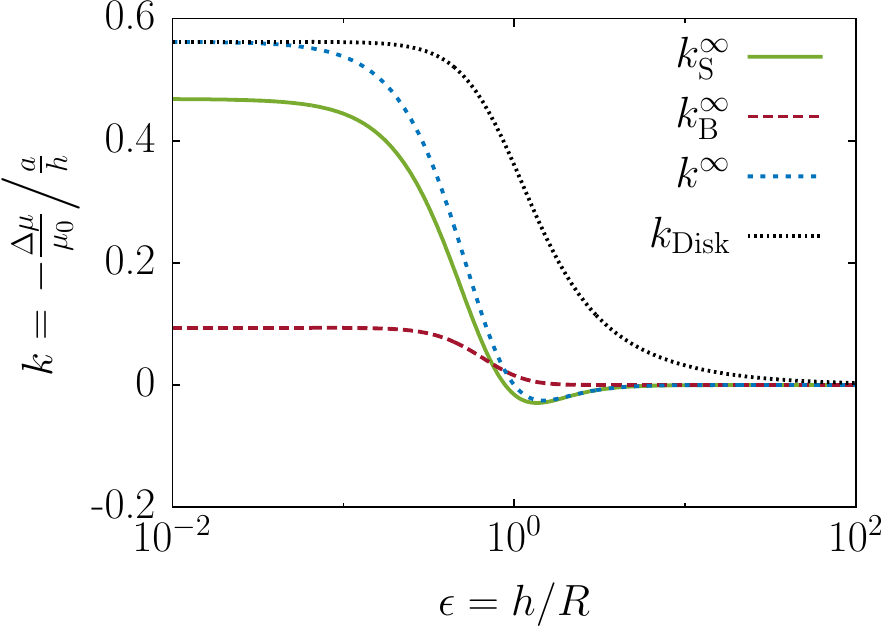}
	\caption{(Color online) Variation of the scaled correction factor to the hydrodynamic mobility function for infinite shear and bending rigidities for the translational parallel motion versus the ratio of the particle-membrane separation to the radius of the membrane. 
	Results are presented for idealized elastic membranes with pure infinite shear rigidity (green solid line), with pure infinite bending rigidity (red long dashed line), as given by Eqs.~\eqref{DeltaMuFinal}, or possessing both infinite shear and bending rigidities (blue short dashed line).
	The corresponding correction factor near a hard no-slip disk as given by Eq.~\eqref{DeltaMuHard} is shown as a black dotted line. 
	Here, we set $B=1$.}
	\label{Fig:Steady-Mobi}
\end{figure}

For completeness, we recall the scaled correction factor to the frequency-dependent hydrodynamic mobility for parallel translational motion near an infinitely extended planar elastic membrane endowed with finite resistance toward shear and bending as
\begin{subequations}\label{CorrectionFaktorsInfMem}
	\begin{align}
		k_\mathrm{S}(\epsilon=0) &= \frac{3}{4} \int_0^\infty
		\frac{\chi (u) \, e^{-2u} }{2Bu^2-\beta^2 + i\beta u \left(2+B\right)}  \, \Intd u \, , \\
		k_\mathrm{B}(\epsilon=0) &= \int_0^\infty
		\frac{3 u^5 \, e^{-2u}}{8u^3 + i \betaB^3} \, \Intd u  \, , 
	\end{align}
\end{subequations}
where
\begin{equation}
	\chi(u) = u \left( Bu \left(u^2-2u+3\right) + i\beta \left( u^2-2u+1+B \right) \right) .
\end{equation}

The convergent improper integrals in Eqs.~\eqref{CorrectionFaktorsInfMem} can be evaluated and expressed in terms of familiar analytic functions as
\begin{subequations}
	\begin{align}
		k_\mathrm{S}(\epsilon=0) &= \frac{15}{32} -\frac{3}{8} \Bigg( \frac{\beta^2}{8}-\frac{3i\beta}{8} + \beta' \, e^{\beta' } \mathrm{E}_1 \left( \beta' \right) \notag \\
		 &\quad+\left( \frac{i\beta}{2} \left(1-\frac{\beta^2}{4} \right) -\frac{\beta^2}{2} \right)e^{i\beta} \mathrm{E}_1(i\beta)
		 \Bigg), \label{kSInf} \\
		 k_\mathrm{B}(\epsilon=0) &= \frac{3}{32} - \frac{i\beta_{\mathrm{B}}^3}{64}\left( \zeta(\betaB) + e^{-i\beta_{\mathrm{B}}}\mathrm{E}_1(-i\beta_{\mathrm{B}}) \right) , \label{kBInf}
	\end{align}
\end{subequations}
with $\beta' = 2i\beta/B$ and 
\begin{equation}
 \zeta(\betaB) =  e^{-i \overline{z_{\mathrm{B}}}} \mathrm{E}_1 \left(-i\overline{z_{\mathrm{B}}} \right) +  e^{-i z_{\mathrm{B}}} \mathrm{E}_1 \left(-iz_{\mathrm{B}} \right) \, , 
\end{equation}
where $z_{\mathrm{B}}=j\beta_\mathrm{B}$ and $j=e^{2i\pi/3}$. 
Here, the bar stands for a complex conjugate and $\mathrm{E}_1$ denotes the exponential integral function of order one~\cite{abramowitz72} defined by $\mathrm{E}_1(x)=\int_1^\infty t^{-1} e^{-xt} \, \mathrm{d} t$.

\begin{figure}
	\includegraphics[scale=1]{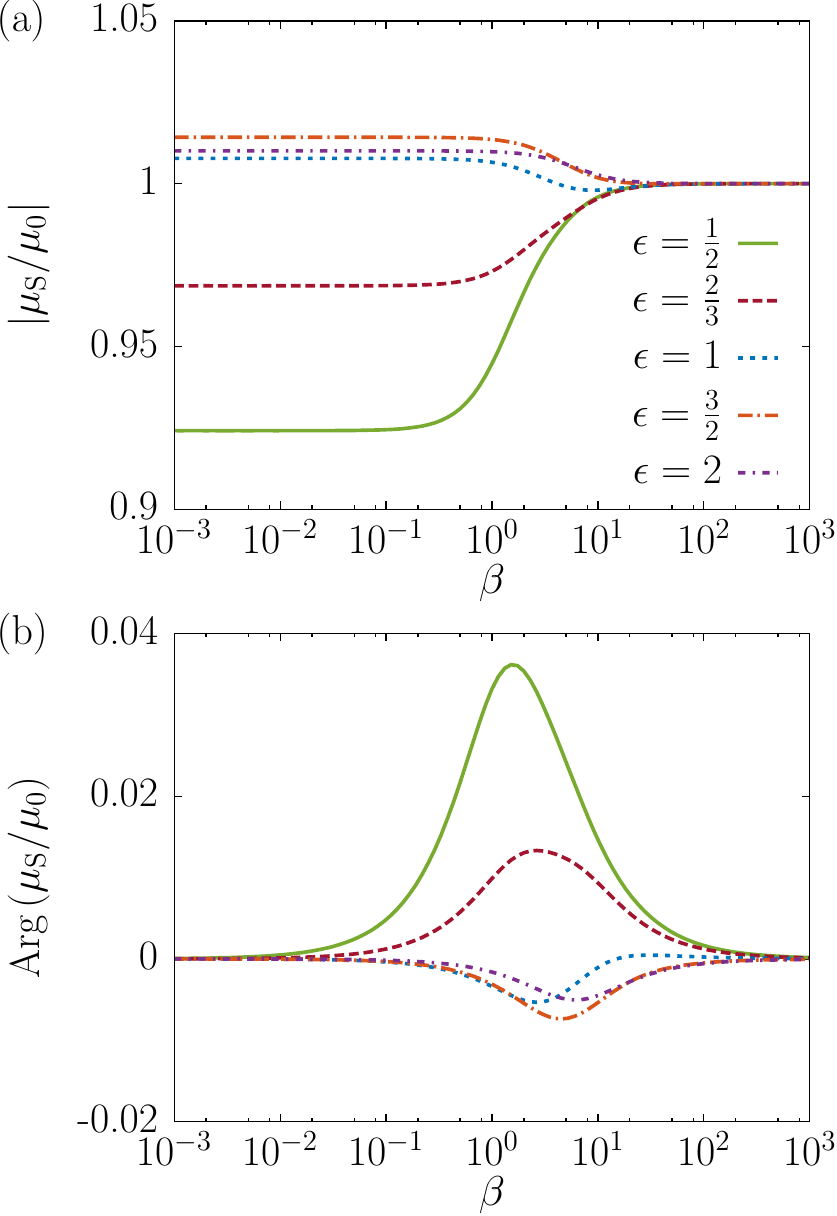}
	\caption{(Color online) (a) Magnitude and (b) phase of the scaled  hydrodynamic mobility function versus the scaled characteristic frequency associated with shear $\beta = 2h/\alpha$ for translational motion parallel to a finite-sized elastic membrane with pure shear resistance.
	Results are shown for various values of the ratio of the particle-membrane separation to the radius of the membrane.
	Here, we set $a/h = 1/2$.}
	\label{Fig:FreqDep-Mobi-Shear}
\end{figure}

\begin{figure}
	\includegraphics[scale=1]{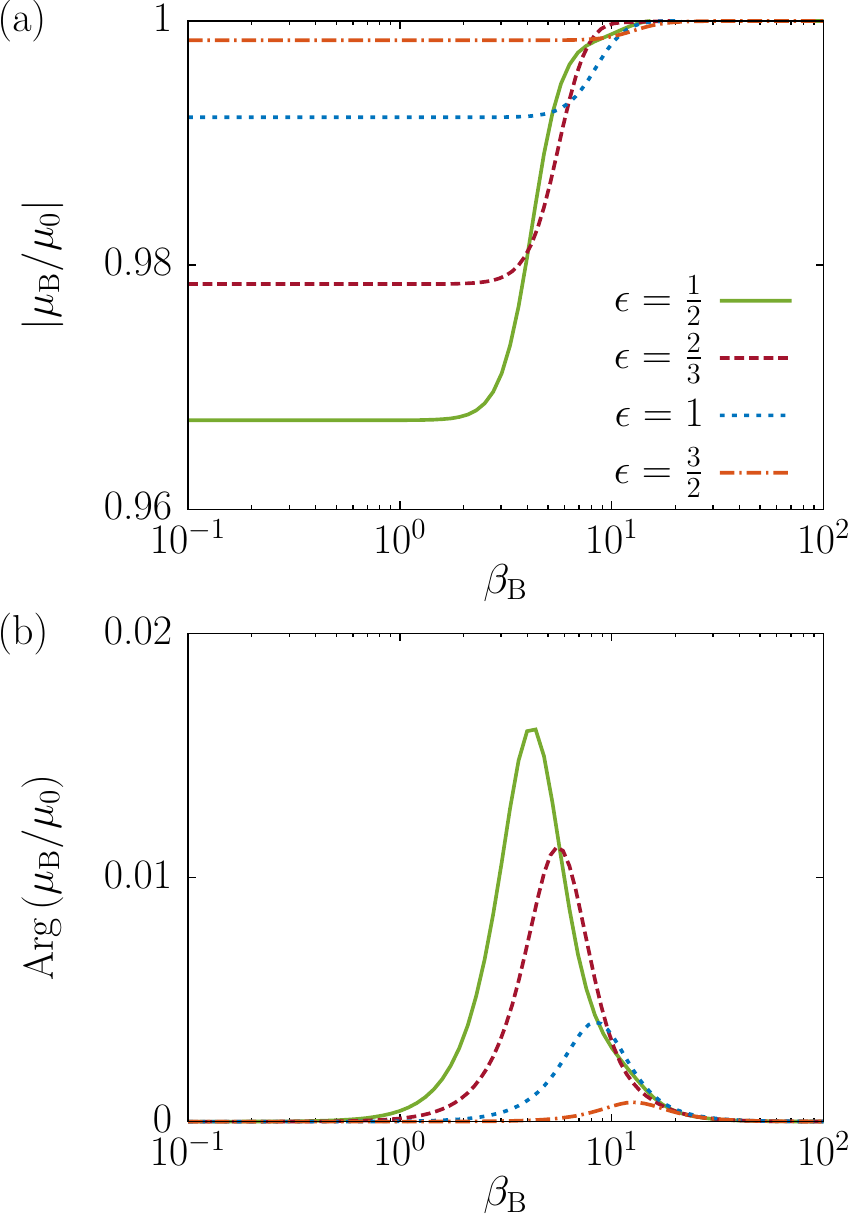}
	\caption{(Color online) (a) Magnitude and (b) phase of the scaled particle mobility versus~$\betaB = 2h/\alphaB$ near a membrane endowed with resistance toward bending for various values of the ratio of the particle-membrane separation to the radius of the membrane.
	Here, we set $a/h = 1/2$.}
	\label{Fig:FreqDep-Mobi-Bending}
\end{figure}


In Fig.~\ref{Fig:Steady-Mobi} we present the variation of the scaled correction factor to the hydrodynamic mobility of a point-like particle located at a distance~$h$ above a finite-sized elastic membrane of radius~$R$.
We display results for idealized membranes of pure infinite shear rigidity (green solid line), of pure infinite bending rigidity (red long dashed line), and of combined infinite shear and bending rigidities (blue short dashed line), see Eqs.~\eqref{DeltaMuFinal} for the corresponding expressions.
Here, we set $B=1$.
For comparison, we also include in the same plot the curve corresponding to the correction factor near a hard no-slip disk (black dotted line) stated by Eq.~\eqref{DeltaMuHard}.

As might intuitively be expected, the correction factor for translational motion parallel to the disk is mainly determined by shear resistance so that bending does not play a dominant role.
This behavior is in stark contrast to the axisymmetric motion perpendicular to the disk where the dynamics has been proven to be mainly dominated by the resistance against bending~\cite{daddi19jpsj}.
We observe that for a membrane of infinite shear rigidity, the correction factor varies non monotonically with~$\epsilon$ and reaches a minimum value at~$\epsilon=1.36$.
Interestingly, the correction factor becomes negative above a threshold values of about $\epsilon_\mathrm{0} \simeq 0.87$, implying that translational motion is sped up when compared to the motion in bulk fluid.
For an elastic membrane simultaneously endowed with both infinite shear and bending rigidities, this threshold value is shifted to a slightly larger value of about 1.00.

For $B>4$, it can be shown that the correction factor near an idealized membrane with pure energetic resistance toward shear takes strictly positive values in the whole range of~$\epsilon$.
However, since physically, the parameter $B := 2/(1+C) \in (0,2]$, a slight increase in mobility always occurs above a certain threshold value~$\epsilon$.
This counterintuitive result might be a direct consequence of the tangential motion of the elastic substrate, inducing an \enquote{effective slippage}. 
Accordingly, the elastic-liquid no-slip condition coupled to a substantial horizontal deformation/displacement of the elastic membrane would induce a large non-zero fluid velocity at the membrane. 
Then, this reduces the hydrodynamic stresses in the gap, and eventually renders the membrane \enquote{invisible} (bulk-like situation).
This surprising effect might further be assessed in future studies for more elaborate model membranes using high-fidelity computer simulations.

For a finite-sized system $(\epsilon \ne 0)$, the correction factor near a membrane endowed simultaneously with infinite shear and bending rigidities is found to be pronouncedly smaller than that predicted near a hard disk of the same size.
Consequently, the hydrodynamic mobility of the particle in the latter case is much lower than in the former.
This behavior can be understood by the fact that motion near a hard no-slip disk is significantly more restricted owing to the imposed zero velocity boundary condition, implying an additional hindrance in particle motion. 
In contrast to that, a finite-sized membrane will necessarily undergo free translational motion.
Therefore, motion of the nearby particle is less impeded.
For $\epsilon \ll 1$, it follows from Eqs.~\eqref{seriesKInf} and \eqref{seriesKHard} that the difference between the scaled correction factors~$k^\infty$ and~$k_\mathrm{Disk}$ decays rapidly as the second power of the radius $R$ of the membrane before it eventually vanishes in the limit of an infinitely extended membrane, for which $k^\infty(\epsilon=0) = k_\mathrm{Disk}$. 
In this limit, the behavior of the particle near a membrane possessing both infinite shear and bending resistances is equivalent to that near a hard disk of no-slip surface conditions.


In Fig.~\ref{Fig:FreqDep-Mobi-Shear}, we present the variation of the amplitude and phase (argument) of the scaled hydrodynamic mobility $\mu_\mathrm{S}/\mu_0 = 1 - k_\mathrm{S} (a/h)$ as a function of the scaled frequency associated with the shear-deformation mode for a membrane of only energetic shear resistance.
Here, we set $a/h = 1/2$.
Results are shown for various values of~$\epsilon$, which span the values to be expected for a wide range of practical situations. 
We observe that the amplitude of the particle mobility displays a sigmoidal, logistic-like phenomenology that amounts to~$1 - k_\mathrm{S}^\infty (a/h)$ for $\beta=0$ and monotonically approaches one as $\beta\to\infty$.
In the latter limit, the system exhibits a bulk-like behavior. 
As already mentioned, for relatively large membranes for which $\epsilon < \epsilon_0$, where $k_\mathrm{S}^\infty (\epsilon_0) = 0$, it follows that $|\mu| < \mu_0$.
Thus, the dynamics of the particle in this case is hindered.
In contrast to that, particle motion is sped up for relatively small membranes for which $\epsilon > \epsilon_0$.
In addition, the argument of the mobility function shows the typical Gaussian bell-shaped curve that peaks at intermediate frequencies around $\beta \sim 1$.
This behavior is a clear signature of dynamic coupling to elastic deformations of the membrane.

In Fig.~\ref{Fig:FreqDep-Mobi-Bending}, we present the corresponding curves for the scaled mobility near an idealized membrane with pure bending resistance $\mu_\mathrm{B}/\mu_0 = 1 - k_\mathrm{B} (a/h)$.
In contrast to a membrane with pure shear resistance, the amplitude of the scaled mobility reaches a minimum value of $1 - k_\mathrm{B}^\infty (a/h)$ in the zero-frequency limit and increases monotonically with frequency to approach one as $\betaB\to\infty$.
In particular, the resulting amplitudes are notably larger than those near an idealized membrane with pure shear resistance.
Therefore, the correction to the particle mobility near a finite-sized membrane is to a bigger degree determined by shear resistance.
Overall, our results suggest that accounting for the finite size of the membrane is important over a large range of frequencies so as to ensure a reliable estimation of the hydrodynamic mobility.

\section{Conclusions}
\label{sec:abschluss}

To summarize, we have presented a fully analytical theory for the asymmetric Stokes flow induced by a transversely directed point-force singularity acting near a finite-sized elastic membrane possessing shear elasticity and/or bending rigidity. 
In conjunction with the results obtained for the axisymmetric flow problem previously treated, the general solution for an arbitrary point-force direction with the point force located on the symmetry axis of the initially undeformed membrane can now be addressed. 
We have formulated the solution of the flow problem in terms of a system of dual integral equations, which we have then reduced into a system of integro-differential equations for unknown wavenumber-dependent functions amenable to numerical integration. 
In addition, we have derived semi-analytical expressions for the mobility function of a point-like particle translating tangent to the finite-sized membrane, showing that the system behavior for infinite shear and bending rigidity of the membrane is mainly dominated by the shear resistance.
Most importantly, we have found that, near an elastic membrane with pure infinite shear resistance, translational motion can be sped up compared to the bulk behavior in some range of parameters.

In this contribution, we have assumed as a first step that the singularity is located on the symmetry axis of the undeformed circular membrane.
The solution of the elastohydrodynamic problem for a point force located at an arbitrary position would be worth investigating in a future work. 
Another possible extension of the present results would be to quantify the possible effect of a hydrodynamic lift force exerted by the membrane on the translating particle using the Lorentz reciprocal theorem of Stokes flow~\cite{rallabandi18}. 
We hope that our results will stimulate additional works on the effect of elastic confinements on the behavior of particulate flows and pave the way toward potential scientific applications in microfluidics and biomedical engineering.

\begin{acknowledgments}
	We would like to thank Stephan Gekle, Maciej Lisicki, Badr Kaoui, Andreas M.\ Menzel, and Hartmut Löwen for early collaboration on this work, and Thomas Salez for fruitful discussions.
	The author gratefully acknowledges support from the DFG (Deutsch Forschungsgemeinschaft) through the project DA~2107/1-1.
\end{acknowledgments}


%

\end{document}